\documentclass{pasj00}

\begin{document}
\SetRunningHead{Y. Katsukawa et al.}{Formation Process of a Light Bridge}
\Received{}
\Accepted{}

\title{Formation Process of a Light Bridge Revealed with 
the Hinode Solar Optical Telescope}

\author{%
  Yukio \textsc{Katsukawa}\altaffilmark{1},
  Takaaki \textsc{Yokoyama}\altaffilmark{2},
  Thomas E. \textsc{Berger}\altaffilmark{3},
  Kiyoshi \textsc{Ichimoto}\altaffilmark{1},
  Masahito \textsc{Kubo}\altaffilmark{4},\\
  Bruce W. \textsc{Lites}\altaffilmark{4},
  Shin'ichi \textsc{Nagata}\altaffilmark{5}
  Toshifumi \textsc{Shimizu}\altaffilmark{6},
  Richard A. \textsc{Shine}\altaffilmark{3},
  Yoshinori \textsc{Suematsu}\altaffilmark{1},\\
  Theodore D. \textsc{Tarbell}\altaffilmark{3},
  Alan M. \textsc{Title}\altaffilmark{3}, and
  Saku \textsc{Tsuneta}\altaffilmark{1}
}
\altaffiltext{1}{National Astronomical Observatory of Japan, 2--21--1 Osawa, 
Mitaka, Tokyo 181--8588}
\altaffiltext{2}{Department of Earth and Planetary Science, University of 
Tokyo,  7--3--1 Hongo, Bunkyo-ku, Tokyo, 
113--0033}
\altaffiltext{3}{Lockheed Martin Solar and Astrophysics Lab, B/252, 3251 
Hanover St., Palo Alto, CA 94304, USA}
\altaffiltext{4}{High Altitude Observatory, National Center for Atmospheric
Research, P. O. Box 3000, Boulder, CO 80307, USA}
\altaffiltext{5}{Hida Observatory, Kyoto University, Takayama, Gifu 506--1314}
\altaffiltext{6}{Institute of Space and Astronautical Science, Japan 
Aerospace Exploration Agency, 3--1--1 Yoshinodai, Sagamihara, Kanagawa 
229-8510}

\email{yukio.katsukawa@nao.ac.jp}


%


\maketitle

\begin{abstract}
The Solar Optical Telescope (SOT) aboard {\it HINODE} successfully 
and continuously observed a formation process of a light bridge in a 
matured sunspot of the NOAA active region 10923 for several days
with high spatial resolution. During its formation, many umbral dots were 
observed emerging from the leading edges of penumbral filaments, and 
intruding into the umbra rapidly. The precursor of the light bridge 
formation was also identified as the relatively slow inward motion of the 
umbral dots which emerged not near the penumbra, but inside the umbra. 
The spectro-polarimeter on SOT provided physical conditions in the 
photosphere around the umbral dots and the light bridges. We found the 
light bridges and the umbral dots had significantly weaker magnetic 
fields associated with upflows relative to the core of the umbra, which
implies that there was hot gas with weak field strength penetrating from 
subphotosphere to near the visible surface inside those structures. There 
needs to be a mechanism to drive the inward motion of the hot gas along 
the light bridges. We suggest that the emergence and the inward motion 
are triggered by a buoyant penumbral flux tube as well as the 
subphotospheric flow crossing the sunspot.
\end{abstract}

\section{Introduction}
A sunspot provides us with a unique site to understand interaction
between very strong magnetic fields and convective flows driven by
subsurface heat transfer. Convective flows are especially important
in breakup and disappearance of a sunspot during its lifetime. But 
the process is still poorly understood. One of the well-known signature
of sunspot breakup is formation of a light bridge (LB) which is a 
lane of relatively bright material dividing an umbra into two parts
\citep{bray1964}. The formation is a result of reestablishment of the 
granular surface as a precursor of the decay of a spot \citep{vazquez1973}.
LBs have been classified based on morphological arrangement and brightness.
A strong LB, which separates umbral cores, is further distinguished as
either penumbral or photospheric according to fine structures observed
within it \citep{sobotka1993, sobotka1994}. A faint (or umbral) LB, which 
is a faint narrow lane within the umbra, most likely consists of a chain 
of umbral dots \citep{muller1979}. The classification is somewhat 
phenomenological, but implies that there should be some sort of 
relationship among LBs, umbral dots, and penumbrae.

It is obvious that gas in a LB must have a temperature higher than 
a surrounding umbra because of its brightness. It is important to know 
magnetic and velocity structures of a LB to understand how the hot gas 
is continuously provided to the LB, otherwise the gas inside the LB 
is cooled down and the LB disappear. Spectrometric observations of 
photospheric Zeeman-sensitive lines indicate that LBs typically have a 
weakened magnetic field strength relative to the nearby umbra with field 
lines inclined from the local vertical \citep{lites1991,rueedi1995,leka1997}. 
Furthermore, it is found that field strengths and inclinations increase 
and decrease with height, respectively, by a detailed analysis of Stokes 
spectra \citep{jurcak2006}, which suggest a canopy-like magnetic structure 
above the LB.  There are no systematic findings considering vertical 
velocities in LBs \citep{leka1997}, but a positive correlation between the 
brightness and upflow velocities is reported by \citet{rimmele1997}, 
which is interpreted as evidence of the hot gas originating from 
subphotospheric convection. These observational results can be explained 
theoretically in terms of a cluster model, where an umbra consists of tight 
bundle of isolated flux tubes separated by field-free columns of hot gas 
\citep{parker1979,spruit2006}. But it is still unknown what triggers 
development of a LB, and what is a role of penumbrae and umbral dots 
in the development.

LBs are also important from the viewpoint of chromospheric and
coronal activities. Observations in H$\alpha$ show surges are 
ejected from a LB in some situations \citep{roy1973,asai2001,bharti2007}.
\citet{berger2003} found constant brightness enhancement over a LB 
in 1600 \AA\ ultraviolet images from the {\it Transition Region and 
Coronal Explorer} ({\it TRACE}), and suggested a steady chromospheric
heat source over a LB. \citet{katsukawa2007} found that formation 
of a LB is spatially and temporally coincident with heating of 
coronal loops seen in {\it TRACE} 171\AA\ images. These observations 
suggest that LBs serve a role not only to dissolve a sunspot, but to 
release or dissipate magnetic energies stored in a sunspot.

The Solar Optical Telescope (SOT, \cite{tsuneta2007}, \cite{suematsu2007}, 
\cite{tarbell2007}, \cite{ichimoto2007}) on the new Japanese spacecraft 
{\it HINODE} \citep{kosugi2007} enables us to observe dynamics and evolution
in the photosphere not only with high spatial resolution (0.2 arcsec) under
seeing-free condition, but with uninterrupted coverage longer than a day 
owing to the sun-synchronous polar orbit.  In this paper, we present a 
successful observation of a formation process of a LB with SOT, which took 
place in a big matured sunspot in NOAA active region 10923 from 13 Nov 2006 
through 17. This observation is the first time ever for us to see evolution 
of a  LB longer than several days with high spatial resolution better than 
1 arcsec. 

\section{Filtergram Observation}

\begin{figure*}[t]
\begin{center}
\FigureFile(499bp,211bp){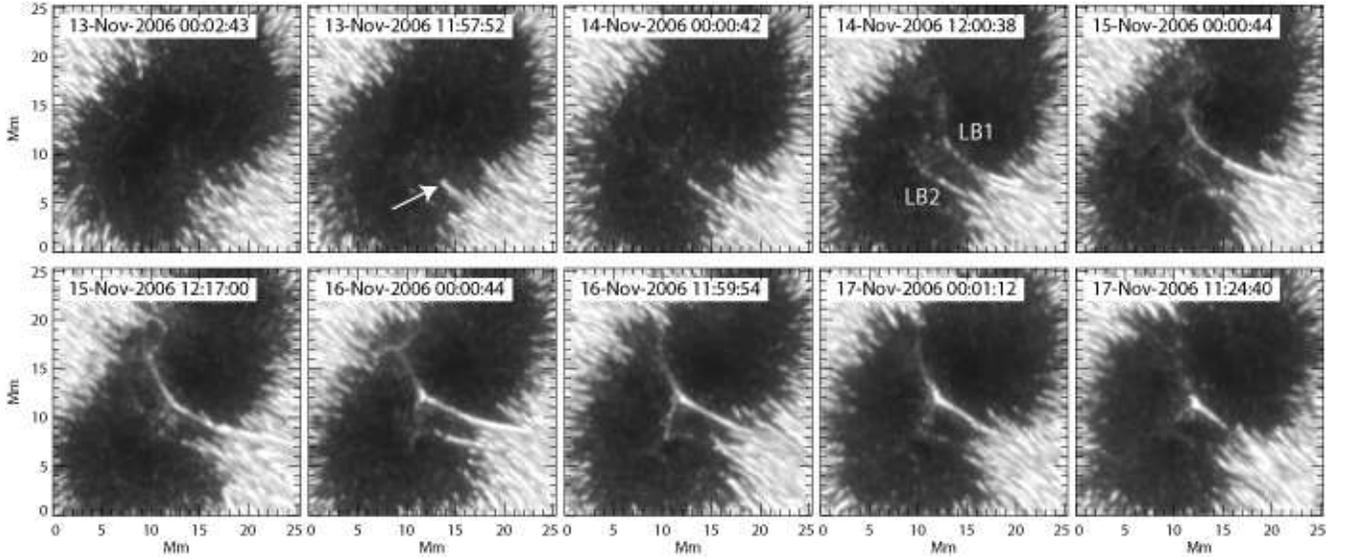}
\caption{Sunspot umbra in NOAA active region 10923 observed with the blue
continuum (BC) channel on SOT from 13-Nov-2006 0UT to 17-Nov-2006 12UT.
Original images covered entire field-of-view of SOT with 2 $\times$2 
summing, but the images here shown are a part of the original images, 
and cover the area of 35'' $\times$ 35'' (25$\times$25 Mm$^2$) centered 
at the sunspot umbra. The arrow in the image at 13-Nov-2006 11:57 indicates
an example of extreme penumbral intrusion observed before formation of 
LBs.}
\label{fig:umbra}
\end{center}
\end{figure*}

We obtained an image sequence of the sunspot continuously through three 
channels G-band 4305 \AA\, Ca II H 3968 \AA\, and blue continuum (BC) 
4504 \AA\ of the broadband filter imager (BFI) on SOT from Nov. 10, 2006 
until the spot went beyond the west limb. The area of 220'' $\times$ 110'', 
which was full field-of-view of BFI, was covered by 2x2 summing (0.108 
arcsec per pixel) in this observation. The time cadence of this BFI 
observation was not so high, and an image through each filter was taken 
every 5 minutes. Here we concentrate on an image sequence taken with the 
BC filter in order to see structures of the lower photosphere inside the 
umbra. Although the continuous observation with BC was interrupted by 
high-cadence observations without BC which lasted for several hours 
from time to time, overall data coverage was excellent for this period, 
especially in terms of monitoring long-term evolution of the LB.

Dark subtraction, flat fielding, correction of bad pixels, and cosmic-ray 
removal are applied for all the filtergram images, and then the observed 
BC intensities are normalized with an average intensity of granules outside 
the sunspot in each image. After that, the filtergram images are spatially 
aligned each other. This process is necessary because drift motion of the 
sunspot is observed in the image sequence because a correlation-tracker 
(CT, \cite{shimizu2007}) on SOT does not track the sunspot, but tracks a 
granule pattern outside the sunspot with a 10'' square field-of-view. 
Intrinsic motion of the sunspot also affects the drift of the sunspot in
the image sequence. The images are aligned each other using the sunspot 
umbra so that centers-of-gravity of the BC intensities in the umbra are 
coincident among the BC images. 

\subsection{Long-term evolution of LBs}

Fig. \ref{fig:umbra} shows time evolution of the sunspot umbra 
observed with BC of BFI from 13-Nov-2006 0UT to 17-Nov-2006 12UT. A movie
of the umbra is available online (see http://www.asj.or.jp/pasj/en/)
which provides dynamical evolution of the umbra and LBs more clearly. 
The sunspot is a big and matured one which has the umbra as dark as 5\% 
of an average intensity of granules outside of the sunspot. There are 
many penumbral filaments surrounding the umbra, and we can see their 
inward migration and intrusion into the umbra everywhere around the 
umbra. Many umbral dots (UDs) are also observed moving inward near the 
boundary of the umbra. Most of them emerge from leading edges of 
the penumbral filaments migrating inward, and become invisible after 
they travel about 2000 km from the leading edges of the penumbral 
filaments. There exist numerous UDs deep inside the umbra, but those are 
generally less distinct than the UDs near the boundary, and their proper 
motions are less significant. These properties of UDs are consistent with 
previous works \citep{kitai1986,grossmann-doerth1986}, in which UDs are 
divided into two classes: peripheral (or penumbral origin) UDs, and 
central (or umbral origin) ones.

The LBs we are now interested in develop from 14-Nov-2006. Even before 
its development, intrusion of penumbral filaments into the umbra is 
outstanding on the southwest side of the umbra where the LBs form after 
that. The similar intrusion of penumbral filaments is seen all around the 
umbra, but the intrusion is extreme in this region because the distance 
of the intrusion is as long as about 4000 km there in some cases. Such 
deep intrusion is preferentially observed around this region during the 
observing period. An example of the intrusion is indicated by the arrow in 
Fig. \ref{fig:umbra}. Another signature of the formation of the LBs is 
numerous central UDs seen in the southwest side of the umbra. Not only 
brightness of each UD but their number density seems larger than the 
rest of the umbra. This is clearly visible in the image at 14-Nov-2006 
00:00 of Fig. \ref{fig:umbra}. 

Internal structures of the umbra drastically changes at around 11:00 
on 14-Nov-2006. UDs born from the penumbral filaments become able to 
migrate deeper than before with much higher velocities on the southwest 
side of the umbra. Many bright UDs are observed to emerge one after 
another, and rush into the umbra from southwest to northeast. The UDs
are called 'rapid UDs' hereafter to distinguish them from the peripheral
or central UDs. The continual emergence and the inward motion of the 
rapid UDs along a trail lead to a chain-like structure of the UDs, which 
can be classified as an umbral LB \citep{muller1979}. There are two LBs 
in which the rapid inward migration of UDs are observed at 12:00 14 Nov. 
The northern one (indicated by LB1 in Fig. \ref{fig:umbra}) is more active
than the southern one (indicated by LB2). The migration of the rapid UDs
lasts for longer than a day along LB1. The bright lane becomes gradually
longer, and finally connects with the opposite side of the umbral 
boundary. The inrush motion of the rapid UDs is investigated in detail
in the next subsection. The inward migration of the rapid UDs is also 
seen along LB2. But the LB does not develop so much, and disappear on 
16 Nov. It needs more investigation to know what makes the difference 
between LB1 and LB2, but this is possibly because the emergence of 
the rapid UDs is less frequent in LB2.

The LB developed along the northern path (LB1) gets brighter after the 
LB connects with the opposite side of the umbral boundary. Especially the 
western part of the LB become bright on 16 Nov. The central part of the 
LB is like a stagnation point of hot gas, and becomes thick and bright. 
After that, bright UDs newly emerge on the southeast side of the central 
bright region. This emergence of UDs from the LB look like leakage of
hot gas from the LB. The motion of the UDs born near the LB are slower 
than that of the peripheral UDs, which is already reported by 
\citet{kitai1986}. The development of the LB stops on 17 Nov. The 
northeastern and southwestern sides of the LB are disconnected from the 
boundary of the umbra, but the central part of the LB stays until 17 Nov. 
After that, the central part of the LB also disappears.

\subsection{Inrush motion of the umbral dots}

\begin{figure*}[t]
\begin{center}
\FigureFile(420bp,290bp){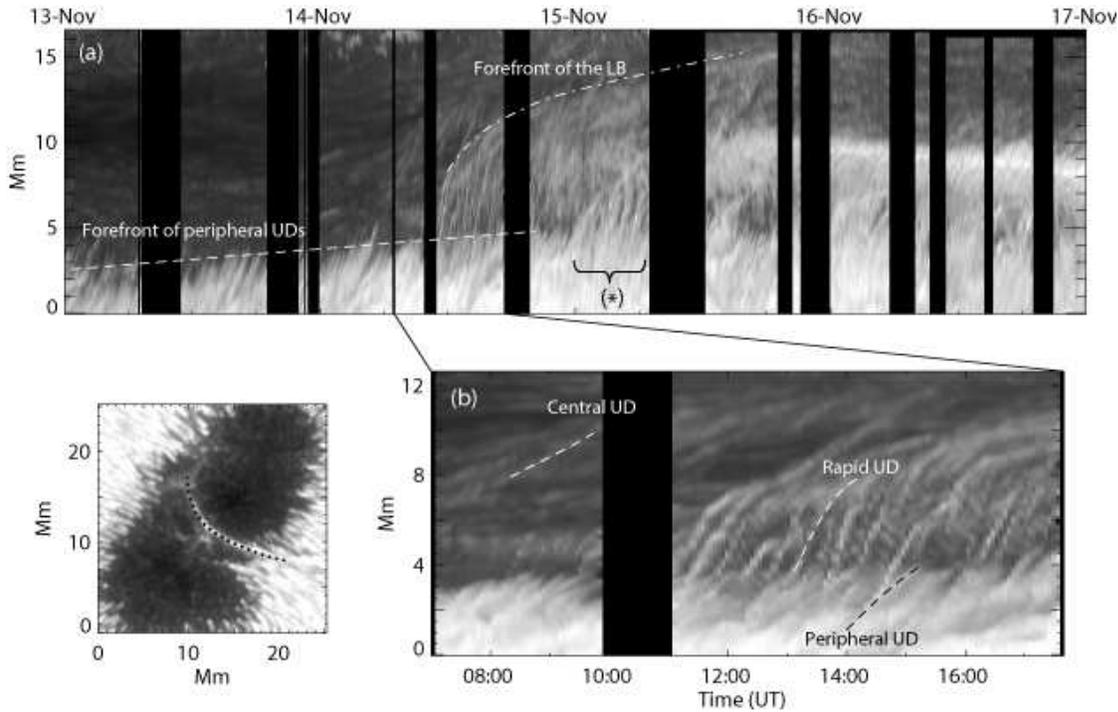}
\caption{Space-time plots along LB1 (dotted curve on the bottom left 
image) from (a) 13-Nov-2006 0:00 to 17-Nov-2006 0:00 and (b) from 7:00 
to 17:30 on 14-Nov-2006 illustrating motion of the UDs and time 
evolution of the LB. The curve along which the space-time plots are created
is not fixed on the images, but motion and change of its shape are taken 
into account. The black zones in the plots represent periods in which BC 
images are not obtained. }
\label{fig:timeslice}
\end{center}
\end{figure*}

Space-time plots shown in Fig. \ref{fig:timeslice} illustrates
time evolution of brightness along LB1. The penumbral side of the
LB has inward motion of penumbral filaments and peripheral UDs for 
all over the period although it is not easy to distinguish the UDs 
with the penumbral filaments in this space-time plots. The velocities 
of the inward motion of the peripheral UDs are about 0.7 km/s, and there 
appear to be no large difference in the speeds among the UDs. The inward 
velocities of the peripheral UDs are almost constant during their travel, 
but they may suffer weak deceleration just before they become invisible, 
which is marginally seen in Fig. \ref{fig:timeslice} (b). The intrusion 
of the peripheral UDs stops at a certain place indicated by a dashed line 
in Fig. \ref{fig:timeslice} (a) for most of the UDs. Some of the 
peripheral UDs have larger intrusion about 2000 km deeper than the 
others. The forefront of the peripheral UDs appears to move inward 
gradually with respect to the center of the umbra before the LB 
formation.

The high speed migration of the rapid UDs starts at around 11:00 14
Nov. The rapid UDs emerge near the leading edge of the penumbral 
filaments quasi-periodically, and rush into the umbra with almost a 
constant velocity until they reach the middle of the umbra. The time 
interval of the emergence is about a quarter of an hour. The inward 
velocities of the rapid UDs are 1 to 2 km/s, which is significantly 
faster than that of the peripheral UDs. The rapid UDs suffer deceleration
near the middle of the umbra, which is clearly seen in Fig. 
\ref{fig:timeslice} (b), but the inward motion lasts with the speed 
slower than 0.5 km/s. Lifetimes of the rapid UDs are mostly several 
tens of minutes, but some of them have a lifetime longer than one hour. 
The continual inward motion of the rapid UDs pushes the forefront of the 
LB gradually as shown by the dash-dotted line in Fig. \ref{fig:timeslice}
(a), and finally the LB reaches the opposite side of the umbral boundary. 
The migration speed of the forefront is about 0.03 km/s. 

\begin{figure*}[t]
\begin{center}
\FigureFile(480bp,100bp){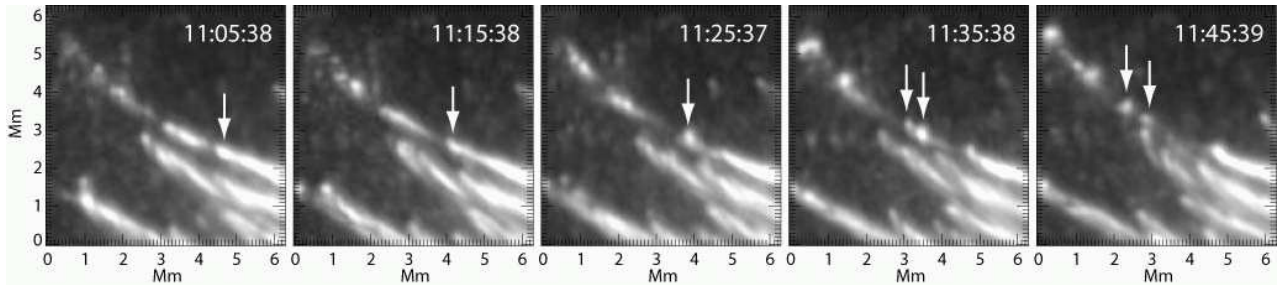}
\caption{Emerging process of rapid UDs from a penumbral filament observed
between 11:05 and 11:45 on 14 Nov. The arrows indicate the rapid UDs we 
are interested in.}
\label{fig:emergence}
\end{center}
\end{figure*}

Fig. \ref{fig:emergence} shows emerging process of rapid UDs at the 
boundary between the umbra and the penumbra. The UDs seem a part of 
a leading edge (or a penumbral grain) of a bright penumbral filament 
before its emergence. The inward migration of the leading edge is 
clearly visible in the BC movie. When the leading edge reaches a 
certain place, elongation and disintegration happens at the leading 
edge of the penumbral filament at around 11:15. The disintegrated part 
becomes a UD, and continues to move inward with the high velocity. 
The UD breaks up into two UDs at 11:35, and both the UDs continue
to migrate deep into the umbra along a channel. The penumbral 
filament stops its inward migration after the emergence of the UD. 

The inward motion of the central UDs are also observed in the deep umbra 
even before the inrush motion of the rapid UDs starts at around 11:00 14 
Nov. An example of the inward motion of the central UDs is indicated by 
the white dashed line in Fig. \ref{fig:timeslice} (b). They are dark and 
less distinct compared with the rapid UDs, but they are long-lived 
(longer than 2 hours). The inward motion of the central UDs begins to 
be observed at around 18:00 13 Nov, which is about 17 hours before the 
inrush motion starts.  At that moment, the inward velocity of the 
central UDs is slower than 0.1 km/s. But the velocity gets larger toward
the formation of the LB, and it is about 0.5 km/s just before the inrush
motion starts. It appears that this inward motion of the central UDs is 
a precursor of the LB formation which is newly identified by HINODE SOT 
thanks to its long-term continuous observation with the high spatial 
resolution.

After the LB develops to some extent, the forefront of the peripheral
UDs changes its evolution at around 0:00 15 Nov. The leading edges 
of the penumbral filaments and the peripheral UDs become able to 
intrude deeper than before. The structures indicated by the asterisks 
in Fig. \ref{fig:timeslice} (a) has brightness similar to the penumbrae, 
and their inward velocities are also similar to those of the penumbrae 
and the peripheral UDs. A measurement of vector magnetic fields in this 
region, which is described in the next section, shows there is a more 
inclined field than the umbra there. It appears that the evolution of 
the LB due to the inrush motion of the rapid UDs is followed by further
intrusion of the penumbral filaments into the umbra, leading to 
formation of the penumbral LB \citep{muller1979}.

\section{Evolution of physical quantities}

\begin{figure*}[t]
\begin{center}
\FigureFile(420bp,450bp){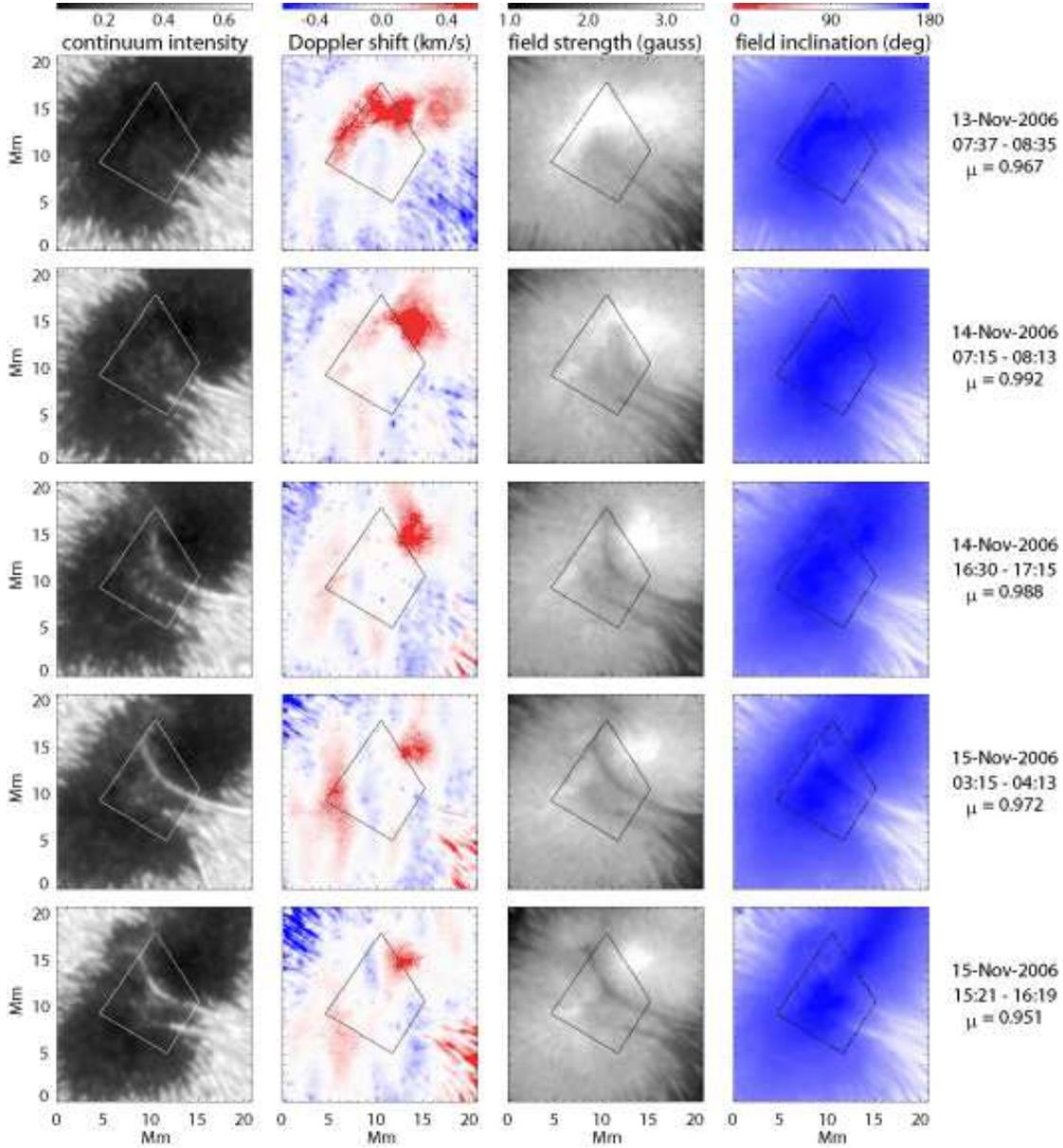}
\caption{Maps of physical parameters obtained by SP. Continuum intensities, 
Doppler shifts, magnetic field strength, and magnetic field inclination 
are shown from left to right. The physical quantities are derived by the 
Milne-Eddignton model fitting to observed Stokes profiles. The Doppler 
shifts are relative to average shifts in the umbra. The positive and 
negative Doppler shifts mean a red and blue shift, respectively. The 
inclination angles are relative to the local normal, which are obtained 
by transformation of field vectors from the observers' reference frame 
with z-axis aligned with the line-of-sight to the local reference frame
with z-axis aligned with the normal to the solar surface. The inclination 
angles of 0 deg, 90 deg, and 180 deg mean that a field vector is outward 
vertical, horizontal, and inward vertical to the surface, respectively. 
The local inclination angles are less affected by the azimuth ambiguity 
because the sunspot is located near the disk center during this observation 
period. The data was taken with the normal mapping mode. The field-of-view 
of the maps here shown is clipped from original maps to present 
20$\times$20 Mm$^2$ around the LB. The  boxes on each map show the area 
used in making the scatter plots in Fig. \ref{fig:scatter}. The rightmost 
column shows observation periods and $\mu$ (cosine of heliocentric 
angles) of the spot for each map.}
\label{fig:spmap}
\end{center}
\end{figure*}

The sunspot was observed also by the spectro-polarimeter (SP) on SOT
during this period. The SP performed scanning observations of the spot 
once or twice per day with the normal mapping mode which scans a target 
with a step size of 0.15 arcsec. An integration time to get spectra is 
4.8 sec at each slit position, and a pixel scale is 0.16 arcsec along 
the slit. The SP observations provide physical quantities in the photosphere
using the two Fe I absorption lines at 6301.5 \AA\ and 6302.5 \AA\ sensitive
to magnetic fields. The quantities described below are derived by 
least-squares fitting to Stokes profiles obtained by SP using the 
Milne-Eddington model atmosphere \citep{yokoyama2007}. Fig. \ref{fig:spmap}
shows two dimensional maps of continuum intensities, Doppler shifts,
magnetic field strength, and magnetic field inclination at the five periods
from 13-Nov-2006 through 15-Nov-2006. Their spatial resolution is high 
and stable enough to spatially resolve LBs and UDs in the maps.

\begin{figure*}[t]
\begin{center}
\FigureFile(440bp,241bp){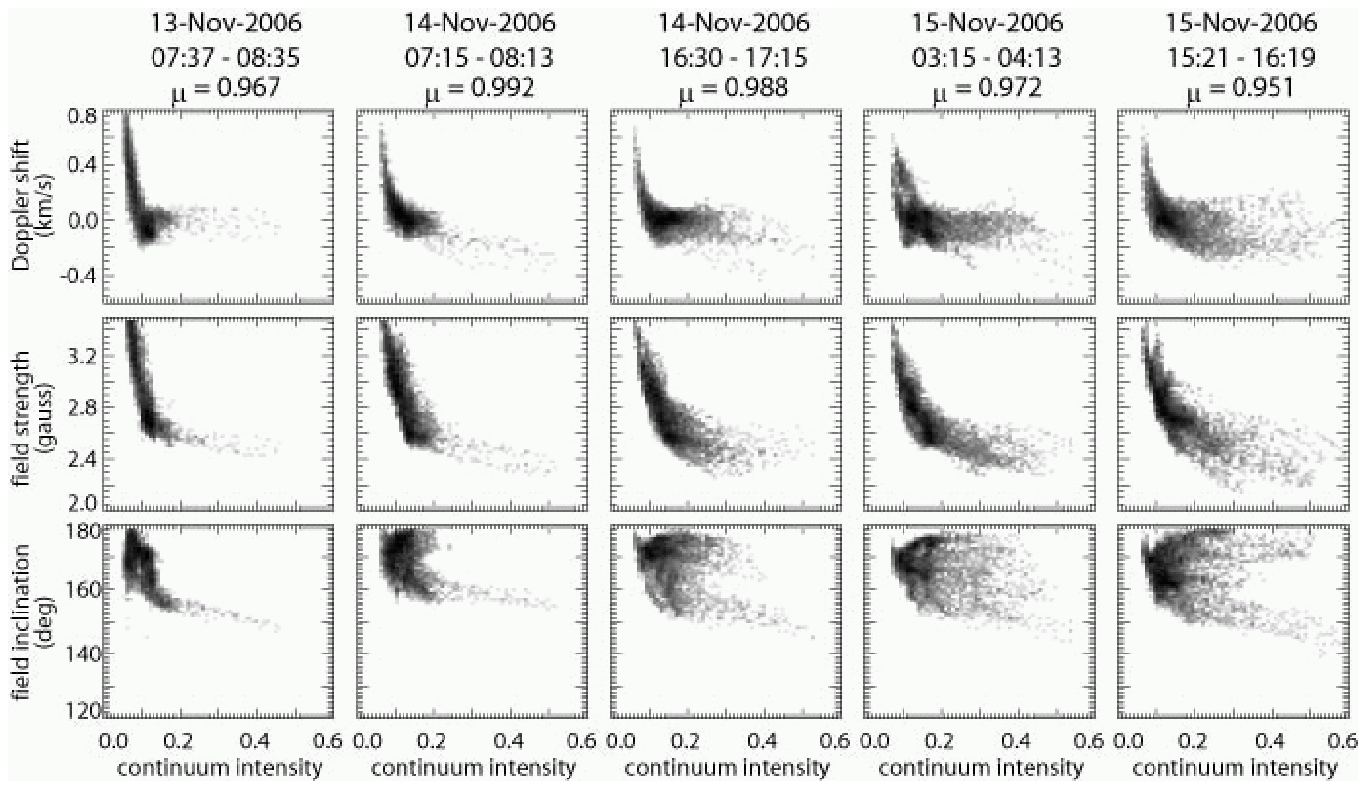}
\caption{Temporal evolution of the Doppler shifts, field strength, and 
field inclination as a function of continuum intensities around the LB.
The area used in the plots is indicated by the boxes on the maps in 
Fig. \ref{fig:spmap}.}
\label{fig:scatter}
\end{center}
\end{figure*}

There is a relatively bright region in the southwest part of the umbra on 
13 Nov, which consists of many central UDs. It is obvious that field 
strength is about 1kG weaker there than the darkest core of the umbra, and 
no significant Doppler shifts. Fig. \ref{fig:scatter} shows correlation 
among the quantities in the photosphere. There is a clear negative 
correlation between the field strength and the continuum intensities inside 
the umbra on 13 Nov. The darkest region in the umbra has a field strength 
about 800 gauss greater than the relatively bright region. This correlation 
was already reported by many authors (see references in \cite{solanki2003}).
A similar correlation is found between the Doppler shifts and the continuum 
intensities. The difference in the Doppler shifts is as much as 0.8 km/s 
between the darkest core and the bright region. Since we do not have an 
absolute wavelength standard in this analysis, it is difficult to determine 
whether the observed Doppler shifts correspond to upflow or downflow with 
respect to the solar surface. If we assume the darkest core of the umbra 
is stationary because gas motion is strongly suppressed by the strong 
magnetic field, the relatively bright region has an upflow, which can be 
interpreted as manifestation of subphotospheric convection below the 
central UDs. 

In the map at 07:15--08:13 14 Nov, which is just before the inward motion
of the rapid UDs appear, some central UDs are visible in the continuum 
intensity map. As we mentioned in the last section, the slow inward motion
of the central UDs are observed at that moment. It is noticed that weak 
field regions in the umbra already have a structure resembling the LBs 
while the structure is not so clear in the Doppler shifts and inclination 
maps. After the rapid UDs appear around 11 UT on 14 Nov, bright structures
in the continuum intensities with weaker field strength evolve along the 
trails. The rapid UDs have magnetic fields a few hundreds gauss weaker 
than the nearby umbra. Most of the rapid UDs have a weak blue shift of 
about 0.2 km/s, but some of the UDs are associated with a slightly larger
blue shift up to 0.5 km/s. These velocities are consistent with a previous 
study using a high spatial resolution spectro-polarimeter 
\citep{socasnavarro2004}. The upflow velocity is as fast as 1.0 km/s if 
we consider the darkest core of the umbra is stationary. The correlation 
between the continuum intensities and the Doppler shifts is not so clear 
near the LBs in Fig. \ref{fig:scatter} contrary to the result obtained by
\citet{rimmele1997}. This means not only the upflows but other factors 
contribute to the brightness of the UDs and the LBs \citep{beckers1977,
spruit2006}. As the LB develops on 15 Nov, further weakening of the field 
strength occurs in the LB. The correlation among the continuum intensities, 
the field strength, and the Doppler shifts becomes less clear as shown in 
the right-most column in Fig. \ref{fig:scatter} although further 
investigation is necessary to understand the reason. 

As for the magnetic field inclination, inclined magnetic fields (30 
degrees from the local vertical) are observed around the leading edge of 
a penumbral filament located on the southwest boundary of the 
region-of-interest in Fig. \ref{fig:spmap} before the LB formation. The 
bright continuum intensities are always associated with the inclined 
fields on 13 Nov, which can be recognized in the leftmost column of 
Fig. \ref{fig:scatter}.  As the LBs develop on 14 -- 15 Nov, bright 
pixels with relatively vertical fields as well as inclined fields grow 
in the region-of-interest. The drastic change of the relation between 
the field inclination and the continuum intensity is easily found in 
Fig. \ref{fig:scatter}. The inclination maps in Fig. \ref{fig:spmap}
shows small contrast around the LBs, which means the inclination in the 
LBs is not so different from the nearby umbra. The extreme intrusion of the 
penumbral filaments along the LBs can be recongnized as less vertical 
inclination angles in the inclination maps in Fig. \ref{fig:spmap}. The 
brightness of the LBs is comparable to that of the leading edge of the 
penumbrae, but there is a significant difference in the inclination. The 
field inclination in LBs is reported to be more inclined than the nearby 
umbra in previous studies (e.g. \cite{leka1997}, \cite{jurcak2006}). We 
cannot find this signature in the LBs we are now interested in. This is 
possibly because the LBs here shown are narrower than in the previous 
studies, and the canopy configuration of magnetic fields might be weak 
in the LBs. Further investigation using many examples of LBs is necessary 
to know general properties of LBs.

\section{Summary and Discussion}

SOT aboard {\it HINODE} provided a precious data set to understand
the formation process of a LB owing to its high spatial resolution
and temporal coverage for several days. We found the following results
from the data set:

1) The LB was resulted from continual emergence of the UDs from the 
leading edges of penumbral filaments and rapid inward migration of 
the UDs. The inward velocity was significantly faster than that of 
peripheral UDs and central UDs.

2) The precursor of the LB formation was identified as the inward
migration of the central UDs. The inward velocity gets faster toward
the LB formation.

3) The extreme intrusion of the penumbral filaments into the umbra
followed the formation of the LB by the rapid UDs. The intrusion made 
a penumbral LB which had magnetic fields inclined to the local vertical.

4) The magnetic field properties and their evolution were revealed 
around the rapid UDs and the LBs with the high and stable spatial 
resolution of the SP. The LBs had significantly weaker magnetic fields 
accompanied by upflows relative to the core of the umbra.

\begin{figure}[t]
\begin{center}
\FigureFile(230bp,200bp){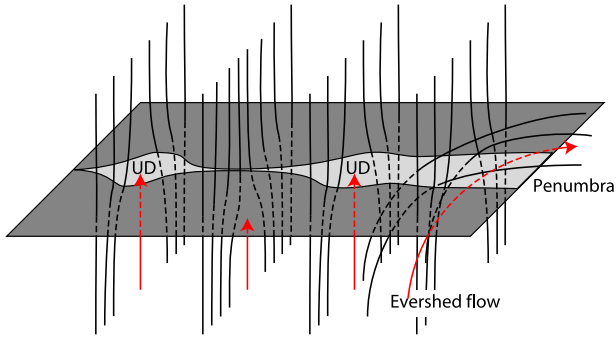}
\caption{Schematic illustration of magnetic field configuration along
a LB near a penumbra. The region colored by dark gray has cold gas with 
strong magnetic fields. The region colored by bright gray has hot gas 
with weak fields, and is accompanied by upflows which are indicated 
by the red arrows.}
\label{fig:model}
\end{center}
\end{figure}

Through this observation, we can get some implication on the breakup 
process of a sunspot. The formation of LBs is a process to breakup 
a sunspot by injection of the hot and weak field gas into the cold and 
strong field gas in the umbra. It is important to know what mechanism 
injects the hot gas into the umbra. The upflows observed in the rapid UDs 
suggests that the hot gas comes from below the photosphere. At the same 
time, the rapid UDs are observed to move inward with the larger velocity.
The inward motion of leading edges of penumbral filaments can be explained 
by a buoyant flux tube \citep{schlichenmaier1998}. On the other hand, the 
inward motion of the rapid UDs may not be associated with the motion of 
field lines, but be the motion of the hot gas pushing through surrounding 
magnetic fields because the magnetic fields inside the umbra got stronger 
if the field lines were transported into the umbra along the inward motion
of the UDs. Each UD has $1.7\times 10^{18}$ Mx magnetic flux when its field 
strength and diameter is 2.5 kG and 300 km, respectively. The total number 
of the rapid UDs moving inward is roughly $\sim 40$ for half a day on 14 
Nov, which provides $\sim 7\times 10^{19}$ Mx magnetic flux into the umbra 
if magnetic fields are transported by the UDs. Increase of field strength 
as high as 70 G should be observed inside the umbra when the transported 
flux was distributed over an area of $10^{18} {\rm cm}^2$. This is 
inconsistent with the observation because gradual decrease of the field 
strength was observed in the umbra during the formation of the LBs. Since 
most of the rapid UDs emerges from the leading edges of the penumbral 
filaments, there should be some driving mechanisms to push the hot gas 
into the umbra near the penumbra.

A possible mechanism is that the emergence of the rapid UDs is related 
to a buoyant flux tube forming the penumbra where a high speed flow, 
i.e. Evershed flow, exists along the tube (see references in 
\cite{solanki2003}). The Evershed flow is an outward directed flow along 
horizontal magnetic fields in the mid penumbra, and an upflow is also 
observed at the inner edge of the penumbra, which may be a source of
the Evershed flow \citep{schlichenmaier2000,rimmele2006}. If the flux 
tube is adjacent to a field free region beneath the photosphere as 
shown in Fig. \ref{fig:model}, the buoyant tube by the hot Evershed
flow may cause enlargement of the field free region, which helps the 
hot gas penetrate to the photosphere by the subphotspheric convection.
The Evershed flow is not stationary, but has significant temporal 
fluctuation with a time scale of 10 to 15 minutes \citep{shine1994, 
rimmele1994}. The emergence of the rapid UDs has the timescale similar 
to the temporal fluctuation of the Evershed flow, which suggests that 
the emergence of UDs might be triggered by high speed blob of the 
Evershed flow. Since it is difficult to prove this mechanism only from 
this observation, we need additional investigation to study the spatial 
and temporal relationship between the Evershed flow and the emergence 
of UDs. 

It is still not clear what makes the difference in the inward velocity
between the rapid UDs and the peripheral UDs. If the Evershed flow is 
the trigger, the rapid motion should be observed everywhere near the 
boundary of the umbra. But the rapid motion is observed only along
the LBs. The most probable reason is that gradual weakening of the 
field strength happens by the emergence and slow inward motion of the 
central UDs before the rapid UDs appear, which was revealed by this 
observation (see Fig. \ref{fig:spmap}). There might be a field free
region beneath the photosphere even before the formation of the LBs,
which helps the intrusion of the rapid UDs into the umbra (see Fig. 
\ref{fig:model}). The intrusion makes further weakening of the field 
strength and helps the catastrophic formation of the LBs. Since the 
central UDs emerged not near the boundary of the umbra, but inside the 
umbra, the Evershed flow cannot explain the inward motion of the central 
UDs. A subsurface flow crossing a sunspot, which is deduced by 
helioseismic studies \citep{zhao2001,kosovichev2002}, may contribute to 
move the central UDs and the formation of the LBs. 

\vspace{\baselineskip}
The authors would like to thank T. Kosugi for his untiring dedication
to the Hinode mission. Hinode is a Japanese mission developed and 
launched by ISAS/JAXA, with NAOJ as domestic partner and NASA and STFC 
(UK) as international partners. It is operated by these agencies in 
co-operation with ESA and NSC (Norway). This work was carried out at the 
NAOJ Hinode Science Center, which is supported by the Grant-in-Aid for 
Creative Scientific Research The Basic Study of Space Weather Prediction 
from MEXT, Japan (Head Investigator: K. Shibata), generous donations 
from Sun Microsystems, and NAOJ internal funding.



\begin{thebibliography}{}

\bibitem[Asai et al.(2001)]{asai2001} Asai, A., Ishii, T.~T., \& 
Kurokawa, H.\ 2001, \apjl, 555, L65
\bibitem[Beckers(1977)]{beckers1977} Beckers, J.~M.\ 1977, \apj, 
213, 900 
\bibitem[Berger \& Berdyugina(2003)]{berger2003} Berger, T.~E., \& 
Berdyugina, S.~V.\ 2003, \apjl, 589, L117 
\bibitem[Bharti et al.(2007)]{bharti2007} Bharti, L., Rimmele, T., 
Jain, R., Jaaffrey, S.~N.~A., \& Smartt, R.~N.\ 2007, \mnras, 376, 1291 
\bibitem[Bray \& Loughhead(1964)]{bray1964} Bray, R.~J., \& 
Loughhead, R.~E.\ 1964, Sunspots, Chapman \& Hall, London
\bibitem[Grossmann-Doerth et al.(1986)]{grossmann-doerth1986} 
Grossmann-Doerth, U., Schmidt, W., \& Schroeter, E.~H.\ 1986, \aap, 156, 347 
\bibitem[Hirzberger et al.(2002)]{hirzberger2002} Hirzberger, J., 
Bonet, J.~A., Sobotka, M., V{\'a}zquez, M., \& Hanslmeier, A.\ 2002, \aap, 
383, 275 
\bibitem[Ichimoto(2007)]{ichimoto2007} Ichimoto, K. et al, 2007, 
\solphys, submitted
\bibitem[Jur{\v c}{\'a}k et al.(2006)]{jurcak2006} Jur{\v c}{\'a}k, J., 
Mart{\'{\i}}nez Pillet, V., \& Sobotka, M.\ 2006, \aap, 453, 1079 
\bibitem[Katsukawa(2007)]{katsukawa2007} Katsukawa, Y. 2007, in ASP 
Conf. Ser., proceedings of the Sixth Solar-B Science Meeting, 
ed. K. Shibata, S. Nagata, \& T. Sakurai, in press
\bibitem[Kitai(1986)]{kitai1986} Kitai, R.\ 1986, \solphys, 104, 287
\bibitem[Kosovichev(2002)]{kosovichev2002} Kosovichev, A.~G.\ 2002, 
AN, 323, 186 
\bibitem[Kosugi(2007)]{kosugi2007} Kosugi, T. et al, 2007, \solphys, in press
\bibitem[Lites et al.(1991)]{lites1991} Lites, B.~W., Bida, 
T.~A., Johannesson, A., \& Scharmer, G.~B.\ 1991, \apj, 373, 683 
\bibitem[Lites et al.(2004)]{lites2004} Lites, B.~W., Scharmer, 
G.~B., Berger, T.~E., \& Title, A.~M.\ 2004, \solphys, 221, 65 
\bibitem[Leka(1997)]{leka1997} Leka, K.~D.\ 1997, \apj, 484, 900 
\bibitem[Muller(1979)]{muller1979} Muller, R.\ 1979, \solphys, 61, 297 
\bibitem[Parker(1979)]{parker1979} Parker, E.~N.\ 1979, \apj, 230, 905 
\bibitem[Rimmele(1994)]{rimmele1994} Rimmele, T.~R.\ 1994, \aap, 
290, 972 
\bibitem[Rimmele(1997)]{rimmele1997} Rimmele, T.~R.\ 1997, \apj, 490, 458
\bibitem[Rimmele \& Marino(2006)]{rimmele2006} Rimmele, T., \& 
Marino, J.\ 2006, \apj, 646, 593 
\bibitem[Roy(1973)]{roy1973} Roy, J.-R.\ 1973, \solphys, 28, 95 
\bibitem[Rueedi et al.(1995)]{rueedi1995} Rueedi, I., Solanki, 
S.~K., \& Livingston, W.\ 1995, \aap, 302, 543 
\bibitem[Schlichenmaier et al.(1998)]{schlichenmaier1998} Schlichenmaier, 
R., Jahn, K., \& Schmidt, H.~U.\ 1998, \aap, 337, 897 
\bibitem[Schlichenmaier \& Schmidt(2000)]{schlichenmaier2000} 
Schlichenmaier, R., \& Schmidt, W.\ 2000, \aap, 358, 1122 
\bibitem[Shine et al.(1994)]{shine1994}
Shine, R.~A., Title, A.~M., Tarbell, T.~D., Smith, K., Frank, Z.~A., 
\& Scharmer, G.\ 1994, \apj, 430, 413 
\bibitem[Socas-Navarro et al.(2004)]{socasnavarro2004} Socas-Navarro, 
H., Pillet, V.~M., Sobotka, M., \& V{\'a}zquez, M.\ 2004, \apj, 614, 448 
\bibitem[Sobotka et al.(1993)]{sobotka1993} Sobotka, M., Bonet, 
J.~A., \& Vazquez, M.\ 1993, \apj, 415, 832 
\bibitem[Sobotka et al.(1994)]{sobotka1994} Sobotka, M., Bonet, 
J.~A., \& Vazquez, M.\ 1994, \apj, 426, 404 
\bibitem[Sobotka et al.(1997)]{sobotka1997} Sobotka, M., Brandt, 
P.~N., \& Simon, G.~W.\ 1997, \aap, 328, 682
\bibitem[Solanki(2003)]{solanki2003} Solanki, S.~K.\ 2003, \aapr, 
11, 153 
\bibitem[Spruit \& Scharmer(2006)]{spruit2006} Spruit, H.~C., \& 
Scharmer, G.~B.\ 2006, \aap, 447, 343 
\bibitem[Shimizu et al.(2007)]{shimizu2007} Shimizu, T. et al., 2007, 
\solphys, submitted
\bibitem[Suematsu et al.(2007)]{suematsu2007} Suematsu, Y. et al., 2007, 
\solphys, submitted
\bibitem[Tarbell et al.(2007)]{tarbell2007} Tarbell, T. D. et al., 2007,
\solphys, in preparation
\bibitem[Tsuneta et al.(2007)]{tsuneta2007} Tsuneta, S. et al., 2007,
\solphys, submitted
\bibitem[Vazquez(1973)]{vazquez1973} Vazquez, M.\ 1973, \solphys, 
31, 377 
\bibitem[Yokoyama et al.(2007)]{yokoyama2007} Yokoyama, T. et al., 
\pasj, 2007, in preparation
\bibitem[Zhao et al.(2001)]{zhao2001} Zhao, J., Kosovichev, 
A.~G., \& Duvall, T.~L., Jr.\ 2001, \apj, 557, 384 
\end{thebibliography}
\end{document}